\documentstyle[12pt,aaspp4,flushrt]{article}
\begin{document}
\title{A LIGHTWEIGHT UNIVERSE?}

\author{
Neta A. Bahcall \& Xiaohui Fan\\
Princeton University Observatory\\
 Princeton, NJ 08544\\
electronic mail : neta,fan@astro.princeton.edu}

\begin{center}
{\em (to appear in {\bf the National Academy of Sciences Proc, 1998)}}
\end{center}

\section*{Abstract}
{\bf How much matter is there in the universe? Does the universe have the
critical density needed to stop its expansion, or is the universe 
underweight and destined to expand forever? We show that several
independent measures, especially those utilizing the largest bound systems
known - clusters of galaxies - all indicate that the mass-density of the
universe is insufficient to halt the expansion. A promising new method, the
evolution of the number density of clusters with time, provides the most
powerful indication so far that the universe has a sub-critical density. We
show that different techniques reveal a consistent picture of a lightweight
universe with only $\sim$ 20-30\% of the critical density. Thus,
the universe may expand forever.}

\vspace{1.5cm}

Standard models of inflation -- how the universe expanded in the 
beginning -- as well as general arguments that demand no
``fine tuning'' of cosmological parameters, predict a flat universe with the
critical density needed to just halt its expansion. 
The critical density, $\rm 1.9 \times 10^{-29} h^{2}g\ cm^{-3}$
(where h refers to Hubble's constant, see below), is equivalent to 
$\sim$ 10  protons
per cubic meter; this density provides the gravitational pull needed to slow down
the universal expansion that began with the Big Bang approximately 15 billion 
years ago, and will eventually bring it to a halt.
So far, however, only a
small fraction of the critical density has been detected, even when all the
unseen ``dark matter'' in galaxy halos and clusters of galaxies is included.
There is no reliable indication so far that most of the matter needed for
closing the universe does in fact exist. Here we show that several
independent observations of clusters of galaxies, including the
mass-to-light ratio of clusters, the high baryon fraction in clusters, and
the observed evolution of cluster abundance, all portray a consistent
picture of a sub-critical universe.

\section{Weighing Clusters} 
Rich clusters of galaxies  -- families of hundreds of galaxies held
together by the gravitational potential of the cluster --
are the most massive bound
objects known.
Cluster masses can be directly and reliably determined using three
independent methods: 1) the motion (velocity
dispersion) of galaxies within clusters reflect the dynamical cluster mass,
within a given radius, assuming the clusters are in hydrostatic
equilibrium (1-3); 2) the temperature of the hot intracluster gas, like the
galaxy motion, traces the cluster mass (4-6); and 3) gravitational lensing
distortions of background galaxies can be used to directly measure the
intervening cluster mass that causes the distortions (7-10). All three
independent methods yield consistent cluster masses (typically within radii
of $\sim$ 1 Mpc $\sim 3 \times 10^{6}$ light years ), indicating that we can reliably determine cluster masses within
the observed scatter ($\sim \pm$ 30\%).

\section{Mass-to-Light Ratio of Clusters} 
Let us begin with the simplest argument for
a low density universe. The masses of rich clusters of galaxies range from   
$\rm \sim 10^{14}\ to\ 10^{15}\ h^{-1}M_{\odot}\ within\ 1.5 h^{-1} Mpc$
radius of the cluster center (where $\rm h =  H_{0}/100\ km\ s^{-1}\ Mpc^{-1}$
denotes Hubble's constant, representing the expansion rate of the
universe). When normalized by the cluster
luminosity, a median mass-to-light ratio of $\rm M/L_{B} \simeq 300 \pm 100 h$ in solar units ($\rm M_{\odot}/L_{\odot}$) is observed
for rich clusters, independent of the cluster luminosity, velocity
dispersion, or other parameters (3,11). 
($\rm L_{B}$ is the total luminosity of the
cluster in the blue band, corrected for internal and Galactic absorption.) 
When integrated over the entire observed luminosity density of the universe,
this mass-to-light ratio
yields a mass density of $\rm \rho_{m} \simeq 0.4 \times 10^{-29} h^{2} g\ cm^{-3}$, or a mass density 
ratio of $\Omega_{m} = \rho_{m}/\rho_{crit} \simeq 0.2 \pm 0.07$
 (where $\rho_{crit}$ is the critical density needed to close the
universe). The inferred density assumes that all galaxies exhibit the same
high M/$\rm L_{B}$ ratio as clusters, and that mass follows light on large scales.
Thus, even if all galaxies have as much mass per unit luminosity as do massive
clusters, the total mass of the universe is only $\sim$ 20\% of the critical density. If
one insists on esthetic grounds that the universe has a critical density
($\Omega_{m}=1$), then most of the mass of the universe has to be unassociated with
galaxies (i.e., with light). On large scales ($\rm \gtrsim 1.5\ h^{-1}\ Mpc$) the mass has to
reside in ``voids'' where there is no light. This would imply, for $\Omega_{m}=1$, a
large bias in the distribution of mass versus light, with mass distributed
considerably more diffusely than light.

Is there a strong bias in the universe, with most of the dark matter
residing on large scales, well beyond galaxies and clusters? A recent
analysis of the mass-to-light ratio of galaxies, groups, and clusters
by Bahcall, Lubin and Dorman (11)
suggests that there is not a large bias. The study shows that the M/$\rm L_{B}$ ratio
of galaxies increases with scale up to radii of $\rm R \sim 0.2\ h^{-1}$ Mpc, due to very
large dark halos around galaxies (see also 12,13). The M/L ratio, however, appears to flatten
and remain approximately constant for groups and rich clusters from scales
of $\sim$ 0.2 to at least $\rm  1.5\ h^{-1}\ Mpc$ and possibly even beyond (Fig.1). The
flattening occurs at $\rm M/L_{B} \simeq 200 - 300 h$, corresponding to $\Omega_{m} \simeq 0.2$. 
(An $\rm M/L_{B}\sim  1350 h$ is needed for a critical density universe, $\Omega_{m}=1$.)
This observation contradicts the classical belief that the relative amount of dark matter
increases continuously with scale, possibly reaching $\Omega_{m}=1$ on large scales. The
available data suggest that most of the dark matter may be associated with
very large dark halos of galaxies and that clusters do not contain a
substantial amount of additional dark matter, other than that associated
with (or torn-off from) the galaxy halos, plus the hot intracluster gas.
This flattening of M/L with scale, if confirmed by further larger-scale
observations, suggests that the relative amount of dark matter does not
increase significantly with scale above $\rm \sim 0.2\ h^{-1}$ Mpc. 
In that case, the
mass density of the universe is low, $\Omega_{m} \sim 0.2 - 0.3$, 
with no significant bias (i.e., mass approximately following light on large scales).

\section{Baryons in Clusters} 
Clusters contain many baryons, observed  as gas and
stars. Within $\rm 1.5 h^{-1}$  Mpc of a rich cluster, the X-ray emitting gas
contributes $\rm \sim 6 h^{-1.5}$ \% of the cluster virial mass (14--16).
Stars
contribute another $\gtrsim$ 4\%. 
The baryon fraction observed in clusters is thus:
\begin{equation}
\Omega_{b}/\Omega_{m} \gtrsim 0.06\rm h^{-1.5} + 0.04
\end{equation}
Standard Big Bang nucleosynthesis limits the baryon
density of the universe to (17--18): 
\begin{equation}
\Omega_{b} \simeq 0.017 \rm h^{-2}
\end{equation}
These facts suggest
that the baryon fraction observed in rich clusters (eq.1)  
exceeds that of an $\Omega_{m} = 1$ universe ($\Omega_{b}/(\Omega_{m}=1) \simeq 0.017\rm h^{-2}$; eq. 2)
by a factor of $\gtrsim$ 3 (for h $\gtrsim 0.5$). 
Since detailed hydrodynamic simulations (14,16)  show that baryons
do not segregate into rich clusters, 
the above results imply that either the mean density
of the universe is lower than the critical density by a factor of $\gtrsim$ 3,
or that the baryon density is much larger than predicted by nucleosynthesis.
The observed high baryonic mass fraction in clusters (eq.1), combined with the
nucleosynthesis limit (eq.2), suggest  (for h $\simeq 0.5 -1$)
\begin{equation}
\Omega_{m} \simeq 0.2 \pm 0.1.
\end{equation}

\section{Evolution of Cluster Abundance} 
In a recent study by  Bahcall, Fan and Cen (19,20) 
we show that the evolution of the number density of clusters as a function of
cosmic time (or redshift) provides a powerful constraint on $\Omega_{m}$ (19-22). 
The growth of high-mass
clusters from initial Gaussian fluctuations depends strongly on the
cosmological parameters $\rm \Omega_{m}\ and\ \sigma_{8}$ 
(where $\sigma_{8}$ is the root-mean-square mass fluctuation on
$\rm 8\ h^{-1}$ Mpc scale; 23--27). 
In low-density models, density fluctuations evolve and
freeze out at early times, thus producing only relatively little evolution
at recent times ($z < 1$). In an $\Omega_{m} = 1$ universe, the fluctuations start
growing more recently thereby producing strong evolution in recent times; a
large increase in the abundance of massive clusters is expected from  $z \sim 1$
to $z \sim 0$. The evolution is so strong in $\Omega_{m}=1$ models that finding even a
few Coma-like clusters at $z > 0.5$ over $\sim 10^{3}\ \rm deg^{2}$ 
of sky contradicts an $\Omega_{m}=1$
 model where only $\sim 10^{-2}$ such clusters would be expected (when normalized
to the observed present-day cluster abundance). The evolution of the number
density of Coma-like clusters was recently determined from observations and
compared with cosmological simulations (19--21). The data show only a slow
evolution of the cluster abundance to $z \sim 0.5$, with $\sim 10^{2}$ times more
clusters observed at these redshifts than expected for $\Omega_{m}=1$. The results
yield $\Omega_{m} \simeq 0.3 \pm 0.1$.

The evolutionary effects increase with cluster mass and with redshift. The
existence of the three most massive clusters observed so far at $ z \sim 0.5-0.9$
places the strongest constraint yet on $\Omega_{m}$ and $\sigma_{8}$. These clusters
(MS0016+016 at $z=0.55$, MS0451+03 at $z=0.54$, and MS1054--03 at $z=0.83$, from
the Extended Medium Sensitivity Survey, EMSS, 28) are nearly twice as massive
as the Coma cluster, and have reliably measured masses (including
gravitational lensing masses, temperatures, and velocity dispersions; 3,9,29--32).
These clusters posses the highest masses ($\rm \gtrsim 8 \times 10^{14}\ h^{-1}\ M_{\odot}$
within 1.5 $\rm h^{-1}$ comoving Mpc radius), 
the highest velocity dispersions ($\rm \gtrsim 1200\ km\ s^{-1}$), and
the highest temperatures ($\rm \gtrsim  8$ kev) in the $z>0.5$ EMSS survey. 
The existence
of these three massive distant clusters, even just the existence of the
single observed cluster at $z=0.83$, rules out Gaussian $\Omega_{m}=1$ models for
which only $\sim 10^{-5}\ z \sim 0.8$ clusters are expected instead of the 1 cluster
observed (or $\sim 10^{-3}\ z > 0.5$ clusters expected instead of the 3 observed).
(See Bahcall \& Fan (29)).

In Figure 2 we compare the observed versus expected evolution of the number
density of such massive clusters. The expected evolution is based on the
Press-Schechter (23) formalism that describes the growth of structure in a
hierarchical universe with standard initial Gaussian density fluctuations;
this formalism agrees well with direct numerical
cosmological simulations (20,26). The expected evolution is shown for different
$\Omega_{m}$ values (each with the appropriate normalization $\sigma_{8}$
that satisfies the
observed present-day cluster abundance, $\sigma_{8} \simeq 0.5 \Omega_{m}^{-0.5}$; 26,33).
The model curves range from $\Omega_{m} = 0.1\ (\sigma_{8} \simeq 1.7$) at the top of the figure (flattest,
nearly no evolution) to $\Omega_{m} = 1\ (\sigma_{8} \simeq 0.5$) at the bottom (steepest,
strongest evolution). The difference between high and low $\Omega_{m}$ models is
dramatic for these high mass clusters: $\Omega_{m}=1$ models predict $\sim 10^{5}$ times
less clusters at $z \sim  0.8$ than do $\Omega_{m} \sim 0.2$ models.
The large magnitude of the effect is due to the fact that these are very massive
clusters, on the exponential tail of the cluster mass function;
they are rare events and the evolution of their number density depends exponentially on 
their ``rarity'', i.e., depends exponentially on $\sigma_{8}^{-2} \propto \Omega_{m}$ (20,23,29).
The number of clusters
observed  at $z \sim 0.8$ is consistent with $\Omega_{m} \sim 0.2$, and is highly inconsistent with the 
$\sim 10^{-5}$ clusters expected if $\Omega_{m}=1$. The data exhibits only a slow, relatively
flat evolution; this is expected only in low --$\Omega_{m}$ models. 
$\Omega_{m}=1$ models
have a $\sim 10^{-5}$ probability of producing the one observed cluster at $z \sim 0.8$,
and, independently, a $\sim 10^{-6}$ probability of producing the two observed
clusters at $z \sim  0.55$. These results rule out $\Omega_{m}=1$ Gaussian models at a
very high confidence level. The results are similar for models with or
without a cosmological constant. The data provide powerful constraints on $\Omega_{m}$
and $\sigma_{8}$: $\Omega_{m}=0.2^{+0.15}_{-0.1}$ and $\sigma_{8} = 1.2 \pm 0.3$ (68\% confidence level) (29).
The high $\sigma_{8}$ value for the mean mass fluctuations 
indicates a nearly unbiased universe, with mass
approximately tracing light on large scales (since the galaxy fluctuations,
which represent the light, exhibit a similar value of $\sigma_{8}$(galaxy) $\simeq 1$).
This conclusion is consistent with the suggested flattening of the observed M/L ratio
on large scales (Fig. 1).

In Figure 3 we summarize the four independent $\Omega_{m}$ ($\sigma_{8}$) constraints obtained
from the cluster results discussed above: 1) the present-day cluster abundance
constraint (26, 33) $\Omega_{m}^{0.5} \simeq 0.5/\sigma_{8}$; 2) the high-redshift ($z \sim 0.5-0.9$) cluster
abundance constraint (29); [the overlap of the $z \sim 0$ and $z \sim  0.5-0.9$ abundance 
constraints of (1) and (2) yields the cluster evolution constraint discussed  above];
 3) the $\Omega_{m}$
derived from the high baryon fraction in clusters; and 4) the $\Omega_{m}$ obtained
from cluster masses. The results are all consistent with each other for 
$\Omega_{m} = 0.2 \pm 0.1$ and $\sigma_{8} = 1.2 \pm 0.2$ (1$\sigma$ level). 
$\Omega_{m}=1$ models are highly
incompatible with these results ($\lesssim 10^{-6}$ probability).

\section{Summary} 
We have shown that several independent observations of clusters of galaxies all
indicate that the mass-density of the universe is sub-critical:
$\Omega_{m} \simeq 0.2 \pm 0.1$.
 A summary of the results, presented in Fig. 3, is highlighted
below.

  1. The mass-to-light ratio of clusters of galaxies and the suggested
     flattening of the mass-to-light ratio on large scales suggest 
     $\Omega_{m} \simeq 0.2 \pm 0.1$.

  2. The high baryon fraction observed in clusters of galaxies suggests 
     $\Omega_{m} \simeq 0.2 \pm 0.1$.

  3. The weak evolution of the observed cluster abundance to $z \sim 1$ provides
     a robust estimate of $\Omega_{m} \simeq 0.2^{+0.15}_{-0.1}$, valid for
     any Gaussian models. An $\Omega_{m}=1$ Gaussian universe is ruled
     out as a $\lesssim 10^{-6}$ probability by the cluster evolution results (Fig.
     2-3).

  4. All the above-described independent measures are consistent with each
     other and indicate a low-density universe with $\Omega_{m} \simeq 0.2 \pm 0.1$ (Fig.3). 
$\Omega_{m}=1$ models are ruled out by the data. While non-Gaussian
     initial fluctuations, if they exist, will affect the cluster evolution
     results, they will not affect arguments (1) and (2) above. 
     Gaussian low-density models (with or without a cosmological constant)
     can consistently explain all the independent observations presented
     here.
These independent cluster observations indicate that we live in a lightweight 
universe with only $\sim 20 \% - 30 \%$ of the critical density, 
Thus, the universe may expand forever.\\

We thank D. Eisenstein, J. P. Ostriker, P. J. E. Peebles, D. N. Schramm, and
D. N. Spergel for helpful discussions. This work was supported in part 
by NSF grant AST93-15368.
\newpage

\begin{center}
 {\bf References}
\end{center}

\vspace*{2mm}
\noindent  1. Zwicky, F. (1957)
 {\em Morphological Astronomy} (Berlin: Springer-Verlag).

\noindent  2. Bahcall, N.A. (1977)
{\em Ann. Rev. Astron. Astrophys.} {\bf 15}, 505 -- 540.

\noindent  3. Carlberg, R.G., Yee, H.K.C., Ellingson, E., Abraham, R., Gravel, P., Morris, S.M., \& Pritachet, C.J. (1996)
 {\em Astrophys.J.} {\bf 462}, 32 -- 49.

\noindent  4. Jones, C. \& Forman, W. (1984)
{\em Astrophys.J.} {\bf 276}, 38 -- 55.

\noindent  5. Sarazin, C.L. (1986)
{\em Rev. Mod. Phys.} {\bf 58}, 1 -- 115.

\noindent  6. Evrards, A.E., Metzler, C.A., \& Navarro, J.F. (1996)
 {\em Astrophys.J.} {\bf 469,} 494 -- 507.

\noindent  7. Tyson, J.A., Wenk, R.A., \& Valdes, F. (1990)
 {\em Astrophys.J.} {\bf 349}, L1 -- L4.

\noindent  8. Kaiser, N. \& Squires, G. (1993)
 {\em Astrophys.J.} {\bf 404}, 441 -- 450.

\noindent  9. Smail, I., Ellis, R.S., Fitchett, M.J., \& Edge, A.C. (1995)
 {\em Mon.Not.Roy.Astron.Soc.} {\bf 273}, 277 -- 294.

\noindent 10. Colley, W.N., Tyson, J.A., \& Turner, E.L, (1996) 
{\em Astrophys.J.} {\bf 461}, L83 -- L86.

\noindent 11. Bahcall, N.A., Lubin, L., \& Dorman, V. (1995)
 {\em Astrophys.J.} {\bf 447}, L81 -- L85.

\noindent 12. Ostriker, J.P., Peebles, P.J.E., \& Yahil, A. (1974)  
{\em Astrophys.J.} {\bf 193}, L1 -- L4.

\noindent 13. Rubin, V.C. (1993) {\em Proc. Natl. Acad. Sci. USA} {\bf 90} 4814 -- 4821

\noindent 14. White, S.D.M., Navarro, J.F., Evrard, A., \& Frenk, C.S. (1993)
{\em Nature} {\bf 366}, 429 - 431.

\noindent 15. White, D. \& Fabian, A. (1995)
{\em Mon.Not.Roy.Astron.Soc.} {\bf 272}, 72 -- 84.

\noindent 16. Lubin, L., Cen, R., Bahcall, N.A., \& Ostriker, J.P. (1996)
 {\em Astrophys.J.} {\bf 460}, 10 -- 15.

\noindent 17. Walker, T.P., et al. (1991)
{\em Astrophys.J.} {\bf 376}, 51 -- 69.

\noindent 18. Tytler, D., Fan, X.-M., Burles, S. (1996)
{\em Nature} {\bf 381}, 207  -- 209.

\noindent 19. Bahcall, N.A., Fan, X., \& Cen, R. (1997)
{\em Astrophys.J.} {\bf 485}, L53 -- L56.

\noindent 20. Fan, X., Bahcall, N.A., \& Cen, R. (1997)
{\em Astrophys.J.} {\bf 490}, L123 -- L126.

\noindent 21. Carlberg, R.G., Morris, S.M., Yee, H.K.C., \& Ellingson, E. (1997)
 {\em Astrophys.J}. {\bf 479}, L19 -- L22.

\noindent 22. Henry, J.P. (1997)
 {\em Astrophys.J}. {\bf 489}, L1 -- L5.

\noindent 23. Press, W.H. \& Schechter, P. (1974)
{\em Astrophys.J}. {\bf 187}, 425 -- 438.

\noindent 24. Peebles, P.J.E. (1993) {\em Principles of Physical Cosmology} (Princeton: Princeton Univ. Press).

\noindent 25. Cen, R. \& Ostriker, J.P. (1994)
 {\em Astrophys.J}. {\bf 429}, 4 -- 21.

\noindent 26. Eke, V.R., Cole, S., \& Frenk, C.S. (1996)
{\em Mon.Not.Roy.Astron.Soc.} {\bf 282}, 263 -- 280.

\noindent 27. Oukbir, J. \& Blanchard, A., (1997)
{\em Astron.Astrophys.} {\bf 317}, 1 -- 13.

\noindent 28. Luppino, G.A. \& Gioia, I.M. (1995)
 {\em Astrophys.J.} {\bf 445}, L77 -- L80.

\noindent 29. Bahcall, N.A. \& Fan, X. (1998)
{\em Astrophys.J}.  in press.

\noindent 30. Luppino, G.A. \& Kaiser, N. (1997)
{\em Astrophys.J.} {\bf 475}, 20 -- 28.

\noindent 31. Mushotsky, R. \& Scharf, C.A. (1997)
{\em Astrophys.J.} {\bf 482}, L13 -- L16.

\noindent 32. Donahue, M., {\em et al.} (1998)
{\em Astrophys.J.} in press.

\noindent 33. Pen, U.-L. (1998)
{\em Astrophys.J.} in press.
\newpage
\begin{figure}
\vspace{-6cm}

\epsfysize=620pt \epsfbox{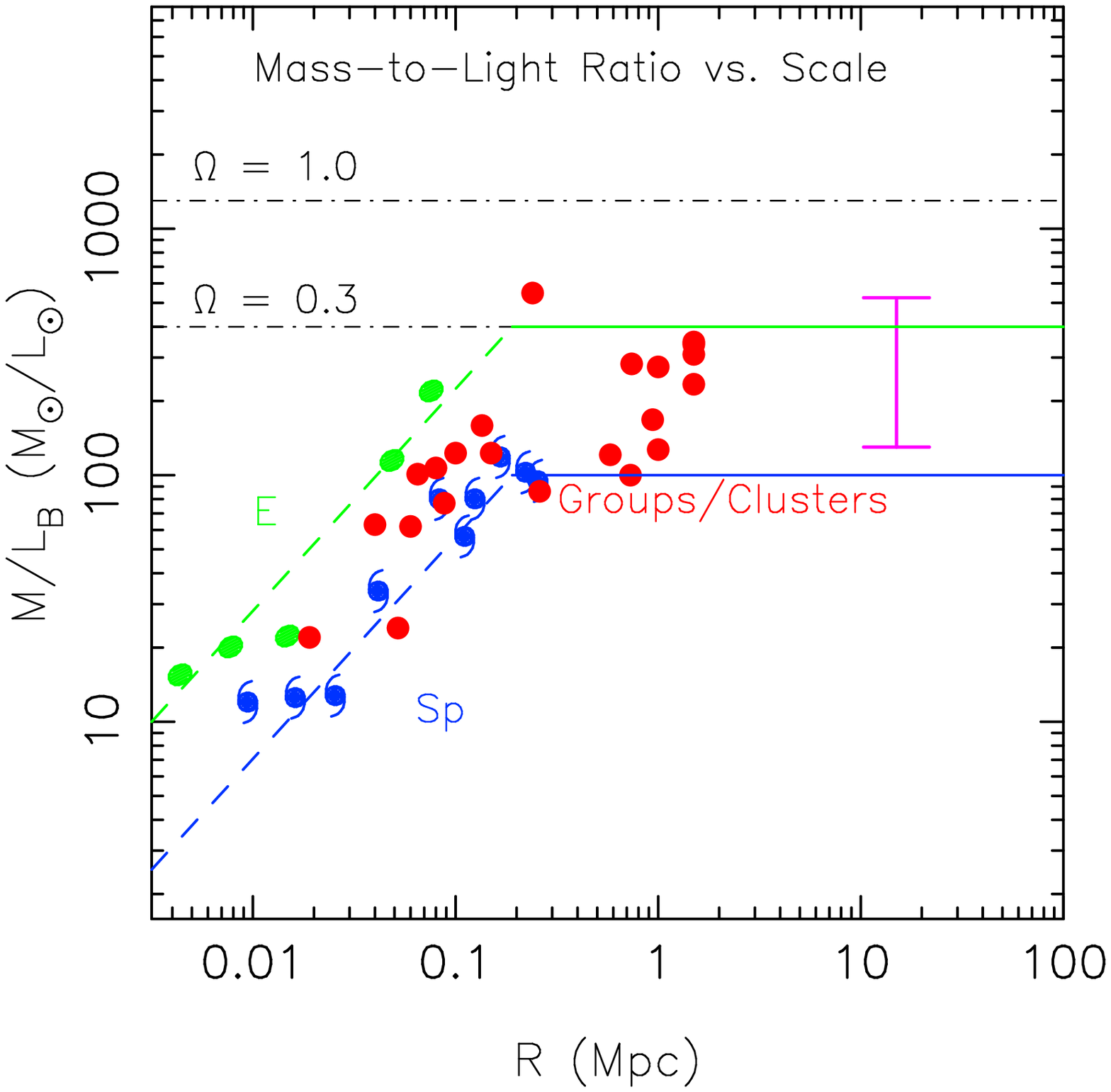}

\vspace{2cm}
\noindent
Figure 1.The dependence of mass-to-light ratio,
$\rm M/L_{B}$, on scale, R, for
average spiral galaxies (blue symbols), elliptical galaxies (green), and
groups and clusters (red).
(From Bahcall, Lubin and Dorman 1995)(11).
The large scale point at $\rm \sim 15 h^{-1}$ Mpc
represents Virgo cluster infall motion results (11).
The location of $\Omega_{m}=1$ and $\Omega_{m}=0.3$ are indicated
by the horizontal lines. A flattening of $\rm M/L_{B}$ is suggested at
$\Omega_{m} \simeq 0.2 \pm 0.1$.
\end{figure}

\begin{figure}
\vspace{-6cm}

\epsfysize=600pt \epsfbox{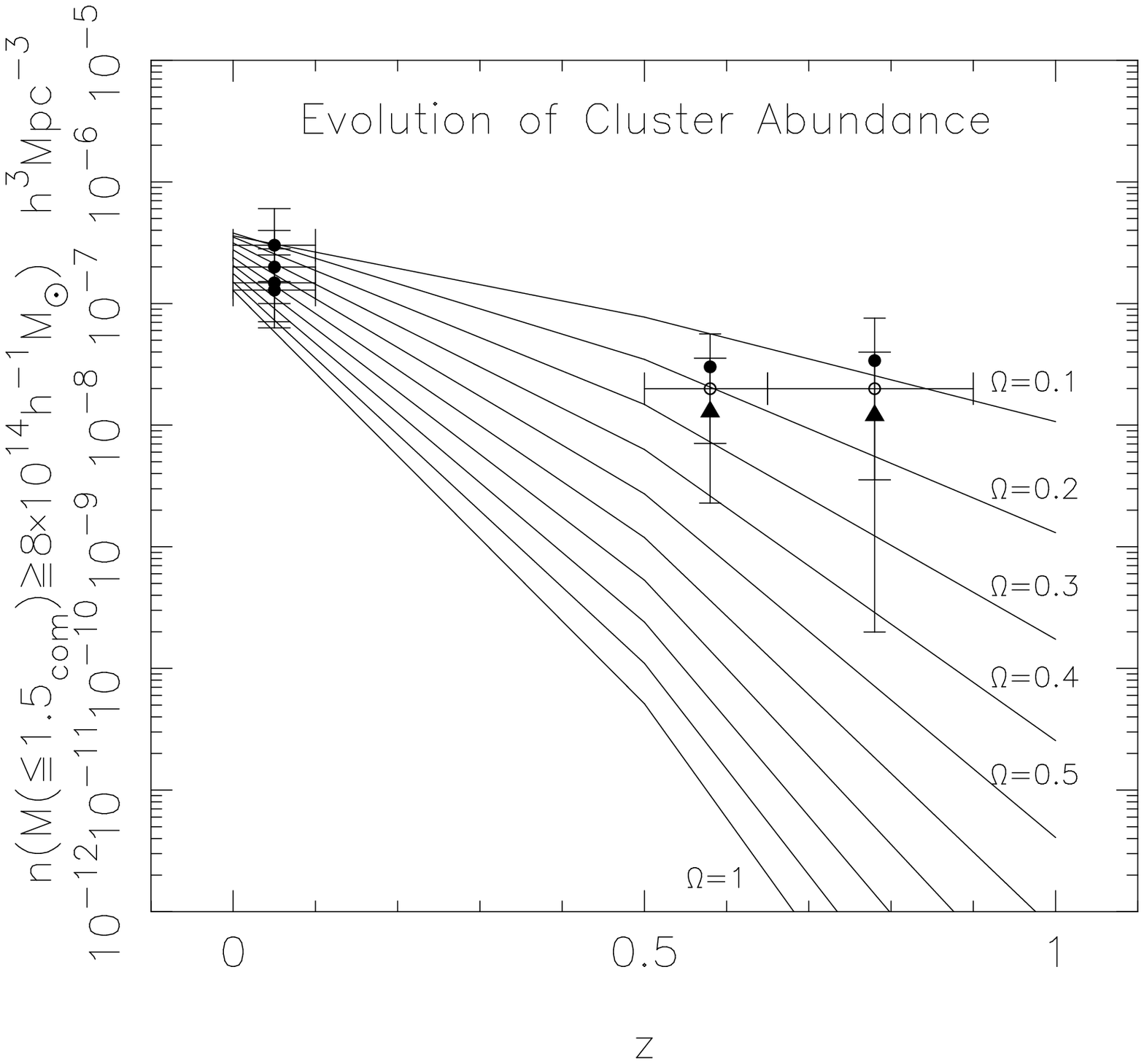}

\vspace{2cm}
\noindent Figure 2. Evolution of the number density of massive
clusters as a function of redshift: observed versus expected
(for clusters with mass $\rm \gtrsim 8 \times 10^{14} h^{-1} M_{\odot}$
within a comoving radius of $\rm 1.5 h^{-1}$ Mpc).
From Bahcall and Fan 1998 (29).
The expected evolution is presented for different
$\Omega_{m}$ values by the different curves.
The observational data points (see text) show only a slow evolution
in the cluster abundance, consistent with $\Omega_{m} \simeq 0.2^{+0.15}_{-0.1}$.
Models with $\Omega_{m} =1 $ predict $\sim 10^{5}$ fewer clusters than observed at $z
 \sim 0.8$, and $\sim 10^{3}$ fewer clusters than observed at $z \sim 0.6$.
\end{figure}
\begin{figure}

\vspace{-3cm}

\epsfysize=600pt \epsfbox{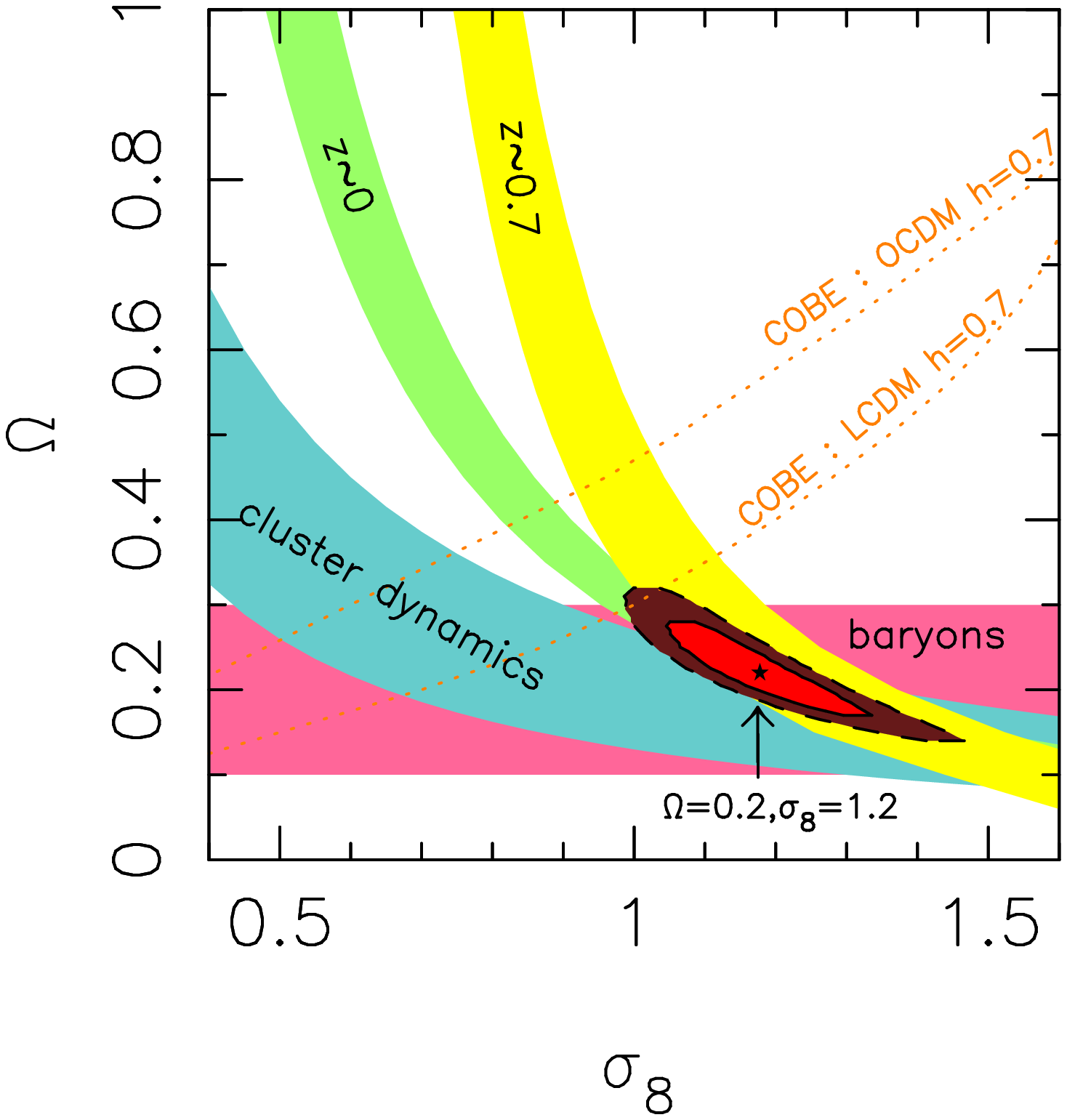}

\vspace{-2cm}
\noindent  Figure 3.  Constraining the mass-density parameter,
$\Omega_{m}$, and the mass fluctuations on $\rm 8 h^{-1}$ Mpc scale, $\sigma_{8}$,
from several independent observations of clusters:
cluster dynamics (blue band); baryon fraction in clusters (pink);
present-day cluster abundance ($z \sim 0 $; green);
and cluster abundance at redshift $z \sim 0.7$ (yellow).
(The latter two abundances yield the cluster
evolution constraints shown in Fig. 2; see text).
 All these model-independent observations converge at the allowed range of
$\Omega_{m} = 0.2 \pm 0.1$ and $\sigma_{8} = 1.2 \pm 0.2$ ( 68\% confidence level; red).
The dotted lines illustrate the mean microwave fluctuations constraints,
based on the COBE satellite results, for
a Cold-Dark-Matter model with h = 0.7 (with and without a cosmological constant, denoted as LCDM and OCDM respectively. Both models are consistent, within their
uncertainties, with the best-fit $\Omega_{m} - \sigma_{8}$ regime of the cluster
observations).
\end{figure}

\end{document}